\documentclass[sts]{imsart}

\RequirePackage[OT1]{fontenc}
\usepackage{amsthm,amsmath,natbib,amssymb,enumerate,mathrsfs}
\RequirePackage[colorlinks,citecolor=blue,urlcolor=blue]{hyperref}

\def\bsigma{{\mbox{\boldmath $\sigma$}}}

\def\bpi{{\mbox{\boldmath $\pi$}}}
\def\bmu{{\mbox{\boldmath $\mu$}}}
\def\btheta{{\mbox{\boldmath $\theta$}}}
\def\bbeta{{\mbox{\boldmath $\beta$}}}
\def\balpha{{\mbox{\boldmath $\alpha$}}}

\def\bx{\textbf{x}}
\def\bI{\textbf{I}}
\def\bX{\textbf{X}}

\def\bw{\textbf{w}}

\def\b0{\textbf{0}}
\def\bh{\textbf{h}}

\def\bu{\textbf{u}}

\startlocaldefs
\numberwithin{equation}{section}
\theoremstyle{plain}

\endlocaldefs
\def\boxit#1{\vbox{\hrule\hbox{\vrule\kern6pt\vbox{\kern6pt#1\kern6pt}\kern6pt\vrule}\hrule}}

\begin{document}

\begin{frontmatter}
\title{An Overview of Semiparametric Extensions of Finite Mixture Models}
\runtitle{Review of semiparametric mixture models}

\begin{aug}
\author{\fnms{Sijia} \snm{Xiang}\ead[label=e1]{sjxiang@zufe.edu.cn}},
\author{\fnms{Weixin} \snm{Yao}\ead[label=e2]{weixin.yao@ucr.edu}}
\and
\author{\fnms{Guangren} \snm{Yang}\ead[label=e3]{tygr@jnu.edu.cn}}

\runauthor{S. Xiang et al.}

\affiliation{Zhejiang University of Finance \& Economics, University of California, Riverside, and Jinan University}

\address{School of Data Sciences, Zhejiang University of Finance \& Economics, Hangzhou, Zhejiang, 310018,
PR China. \printead{e1}.}
\address{Department of Statistics, University of California, Riverside, CA. \printead{e2}.}
\address{Department of Statistics, School of Economics, Jinan University, Guangzhou, China, 510632. \printead{e3}.}
\end{aug}

\begin{abstract}
Finite mixture models have been a very important tool for exploring complex data structures in many scientific areas, for example, economics, epidemiology, finance. In the past decade, semiparametric techniques have been popularly introduced into traditional finite mixture models, and so semiparametric mixture models have experienced exciting development in methodologies, theories and applications. In this article, we provide a selective overview of newly-developed semiparametric mixture models, discuss their estimation methodologies, theoretical properties if applied, and some open questions. Recent developments and some open questions are also discussed.
\end{abstract}

\begin{keyword}
\kwd{EM algorithm}
\kwd{Mixture models}
\kwd{Mixture regression models}
\kwd{Semiparametric mixture models}
\end{keyword}

\end{frontmatter}

\section{Why semiparametric mixture models?}

Parametric mixture models are easy to interpret, fast to estimate, whose theoretical properties have been well studied, and so have been quite popularly used. However, as any other parametric statistical inference, parametric mixture models are all based on some strong model assumptions, taking linearity and normality as examples, and some of the assumptions are often unrealistic in practice. In addition, model mis-specification could be disastrous in parametric mixture models and sometime might lead to misleading results and inference. Please refer to Pommeret and Vandekerkhove (2018) for advantage of a semiparametric method to test a parametric assumption on the unknown component of the two component mixture model with on known component.

To solve this problem, many semiparametric mixture models are proposed to relax assumptions of traditional fully parametric mixture models. Bordes et al. (2016b), Bordes et al. (2007) and Hunter et al. (2007), among others, studied a two-component mixture of locations model where the component density is only assumed to be symmetric. Different estimation methodologies were proposed, but the theoretical properties of which were either studied by later researchers or still remain unattended. Chang and Walther (2007) proposed a mixture of log-concave distributions for clustering. This generalization is general enough to include most of the parametric distribution functions, but non-identifiability was the price to pay for such generality. In addition, a two-component mixture of locations model with a known component has been extensively studied during the past decade, by Bordes et al. (2006a), Bordes and Vandekerkhove (2010), Patra and Sen (2016), Hohmann and
Holzmann (2013), Xiang et al. (2014), Ma and Yao (2015), Huang et al. (2018), and so on. The model is well motivated and various kinds of estimation methods have been studied. However, some of the methods are suffering from not having efficient algorithms or not being able to show the theoretical properties.

In addition, a lot of contributions have been made to relax the parametric assumptions of finite mixtures of regressions (FMR) models. By allowing the mixing proportions to depend on a covariate, Young and Hunter (2010) and Huang and Yao (2012) studied semiparametric mixture of regressions models with varying proportions. Huang et al. (2013) and Xiang and Yao (2016) relaxed the parametric assumptions on the mean functions and/or variances to accommodate for complicated data structure. However, due to the application of kernel regression in the estimation procedure, the models were not suitable for data with high dimensional predictors. Hunter and Young (2012) studied a FMR model where linearity was still assumed within each component, but the error terms were modeled fully nonparametrically. However, since the degrees of freedoms of the aforementioned models are hard to define, the issue of selecting the number of components is still an open question.


The desire for semiparametric mixture models is indeed coming from the practice. 
For example, in order to detect differentially expressed genes under two or more conditions in microarray data, Bordes et al. (2006a) proposed a semiparametric two-component mixture model (\ref{e1.2}), in which one component is known. Practically a test statistic is built for each gene, which, under the null hypothesis, should have a known distribution $f_0$, and have a unknown distribution $f$, otherwise. Then, the collected sample should come from a two-component mixture model with $f_0$ and $f$ as its component distributions. In economics as well, in order to analyze the relationship between the HPI change and the GDP growth, since the scatter plot of the data shows different patterns in different macroeconomic cycles, and the relationship is clearly not linear, Huang et al. (2013) proposed a semiparametric mixture of regressions model (\ref{e6.1}). In practice, since firms tend to manipulate earnings to turn a small loss into a small profit (see for example, Burgstahler and Dichev, 1997 and Ding et al., 2007, which showed that there was strong evidence of manipulations to dramatically boost earnings), Return On Equity (ROE) is actually comprised of real earnings and manipulated earnings. Take this into account, in order to model ROE, Huang et al. (2018) studied the special two-component model (\ref{e1.2}), where the known component $f_0$ is assumed to be Pareto.

Both theoretically and practically, many semiparametric mixture models have been developed and demonstrated to have successful applications during the last few years. In Section 2, we will respectively present a systematic overview of semiparametric mixture of \emph{location} models when Section 3 will be dedicated more specifically to semiparametric mixture of \emph{regression} models. In order to be consistent throughout, we will do our best to use the same notation system, which might not be the same as the original articles. A discussion section ends the paper.

\section{Mixture of locations}
\subsection{Introduction}

Consider a $C$-component mixture model
\begin{equation}
g(\bx)=\sum_{c=1}^C \pi_c f_c(\bx), \hskip 0.2in  \bx\in \mathbb{R}^d,
\label{mmixture}
\end{equation}
where the $f_c$'s are the unknown component densities and $\bpi = (\pi_1, \ldots, \pi_C)^\top$ is a vector of unknown mixture proportions satisfying $\pi_c>0$ for all $c$ and $\sum_{c=1}^C \pi_c=1$. When $C$ is unknown, the selection of $C$ could lead to different convergence rate of the maximum likelihood estimators (MLE) and therefore is a crucial topic, see for example, Leroux (1992), Dacunha-Castelle and Gassiat (1999), and Lemdani and Pons (1999) for different model selection results in the parametric setup.

When the unknown component densities are modeled nonparametrically, (\ref{mmixture}) is referred to by Bordes et al. (2007) and Benaglia et al. (2009) as a semiparametric mixture model. Finite mixture models with nonparametric components are very flexible but these types of models are generally not identifiable without additional restrictions. Hall and Zhou (2003) showed, under some technical conditions, the identifiability of model (\ref{mmixture}) when $C=2$, $d\geq 3$, and $f_c(\bx)$ is expressed as a product of $d$ component-specific marginal density of $\bx$.


\subsection{$d=1$, semiparametric location-shifted mixture model}


When $d=1$, one may impose some shape restrictions on $f_c$, such as, symmetry. Let $f_c(x)=f(x-\mu_c)$ and $f$ be symmetric about the origin, then mixture model (\ref{mmixture}) becomes
\begin{equation}
g(x)=\sum_{c=1}^C \pi_c f(x-\mu_c), \hskip 0.2in  x\in \mathbb{R}.
\label{mmixsym}
\end{equation}
Denote $\btheta=(\pi_1,...,\pi_{C},\mu_1,...,\mu_C)^\top$. Theoretical studies by Bordes et al. (2006b) and Hunter et al. (2007) showed that (\ref{mmixsym}) is identifiable for $C\leq 3$ under some conditions. Specifically, when $C=2, \pi\notin\{0,1/2,1\}$ and $\mu_1\neq \mu_2$, the identifiability holds for the following two-component location-shifted mixture model:
\begin{equation}
g(x) = \pi f(x-\mu_1) + (1 - \pi)f(x-\mu_2), \hskip 0.2in  x\in \mathbb{R},
\label{e1.4}
\end{equation}
where $f(\cdot)$ is zero-symmetric.

Assuming the component distribution to be symmetric, Bordes et al. (2006b) proposed a cumulative distribution function (cdf) based M-estimation method to estimate the Euclidean and functional parts separately, and proved that their estimators are $n^{-1/4+\alpha}$ a.s. consistent for all $\alpha>0$. To be more specific, let $F(\cdot)$ and $G(\cdot)$ be the cdf's of $f(\cdot)$ and $g(\cdot)$. Define $A_\theta=\pi\tau_{\mu_1}+(1-\pi)\tau_{\mu_2}$ (von Neumann, 1931) with $\tau_\mu$ $(\mu\in\mathbb{R})$ being an invertible operator from $L_1$ to $L_1$, then the cdf version of (\ref{e1.4}) is equivalent to $G=A_\theta F$. Let $S_r$ be a symmetry operator defined by $S_rF(\cdot)=1-F(-\cdot)$, then condition $G=A_\theta S_rA_\theta^{-1}G$ happens if and only if $\btheta= \btheta_0$ where $\btheta_0$ is the true value of the parameter, which is in line with the identifiability result listed above.

Define the following divergence function
\begin{equation*}
K(\btheta)=K(\btheta;G)=\int_\mathbb{R}\left\{G_\theta(X)-G(x)\right\}^2dG(x),
\end{equation*}
where $G_\theta=A_\theta S_rA_\theta^{-1} G$. Then, Bordes et al. (2006b) proposed a minimum contrast estimator for $\btheta$, defined by $\arg\min _{\theta\in\Theta} K(\btheta;\hat{G}_n)$, where $\Theta$ is a compact parametric space and $\hat{G}_n$ is the empirical cdf of the sample $(X_1,\ldots,X_n)$ drawn from $G_{\theta_0}$. $F$ is then estimated by $\hat{F}_n=\frac{1}{2}(I+S_r)A_{\hat{\theta}_n}^{-1}\hat{G}_n$, where $I$ is the identity operator, and $I+S_r$ is imposed to guarantee the symmetry of $F$.

Bordes et al. (2007) pointed out that the direct estimator of $f$ in Bordes et al. (2006b) is generally not a probability function (pdf) and the numerical calculation is time consuming. On the other hand, Bordes et al. (2007) proposed to estimate $f$ in (\ref{mmixsym}) by $f_h(x)=\frac{1}{2n}\sum_{i=1}^n\sum_{c=1}^C p_{ic}\{K_h(x-x_i+\mu_c)+K_h(x+x_i-\mu_c)\}$, obtained in an EM context of Bordes et al. (2007), where $p_{ic}$ is the probability that $x_i$ comes from component $c$, $K_h=K(x/h)/h$ and $K(\cdot)$ is a zero-symmetric kernel density function. A generalization of the EM algorithm to model (\ref{mmixsym}) is mainly proposed. However, obtaining the asymptotic behavior of these estimators remains to be an open and challenging task.

Define $d_n(\btheta;\hat{G}_n)=\mathcal{D}[\sum_{c=1}^C \pi_c \hat{G}_n(x+\mu_c), \sum_{c=1}^C \pi_c \{1-\hat{G}_n(x-\mu_c)\}]$ where $\mathcal{D}\{G_1,G_2\}$ is some measure of distance between distributions $G_1$ and $G_2$. Then, Hunter et al. (2007) proposed to estimate $\btheta$ by minimizing $d_n(\btheta;\hat{G}_n)$. They proved that for $C=2$ or $3$, under some technical conditions, the Euclidean part of their estimators is asymptotically normally distributed at the $\sqrt{n}$-rate. Balabdaoui (2017) formally proved the existence of the estimator proposed by Hunter et al. (2007) and established the asymptotic distribution of the estimator.

Butucea and Vandekerkhove (2014) applied Fourier analysis to invert the mixture operator, and related the symmetry of $f$ to the fact that its Fourier transform had no imaginary part. Define $f^*(u)=\int_\mathbb{R}e^{ixu}f(x)dx$ as the Fourier transform of $f(x)$, and denote $M(\btheta,u)=\pi e^{iu\mu_1}+(1-\pi)e^{iu\mu_2}$. Then, model (\ref{mmixsym}) implies
\begin{equation}
g^*(u)=\{\pi e^{iu\mu_1}+(1-\pi)e^{iu\mu_2}\}f^*(u)=M(\btheta,u)f^*(u).
\label{char}
\end{equation}
The symmetry of $f$ implies ${\rm Im}\{g^*(u)/M(\btheta,u)\}=0$ if and only if $\btheta=\btheta_0$. By building a contrast function based on the characteristic function (\ref{char}), the parameter $\btheta$ is then  estimated by $\arg\min_{\theta\in\Theta} S_n(\theta)$ where $S_n(\theta)$ as a U-statistic is an estimator of the contrast $S(\theta)=\int_\mathbb{R}\left\{g^*(u)/M(\theta,u)\right\}^2dW(u)$ with $W$ being a Lebesgue absolutely continuous probability measure supported by $\mathbb{R}$.
Under simpler conditions than Hunter et al. (2007), Butucea and Vandekerkhove (2014) proved the central limit theorem of the estimators, and showed the minimax rates for estimating $f$ to be $n^{-2\beta/(2\beta+1)}$ for some $\beta>1/2$. The authors argued the validity of their estimators and theoretical results for $C\geq 3$ cases, and the identifiability can be verified.

Chee and Wang (2013) proposed a semiparametric MLE approach for the model parameters. Specifically, they suggested to model the unknown density $f$ of (\ref{e1.4}) as
\begin{equation}
\tilde{f}_h(x;Q)=\frac{1}{2}\int\{K_h(x-\sigma)+K_h(x+\sigma)\}dQ(\sigma),
\label{f1}
\end{equation}
where $Q$ is a mixing distribution completely unspecified, as a generalization of the kernel-based method. Even with fixed $\btheta$, the estimation of $Q$ is not a simple task since it is an optimization problem over an infinite dimensional space. Lindsay (1983) showed the existence and uniqueness of the NPMLE of $Q$ and also showed that the NPMLE of $Q$ must be discrete with finite support points no more than the number of observations. Then, let
\begin{equation}
\hat{Q}_n=\sum_{j=1}^m w_j\delta_{\sigma_j}
\label{e2.17}
\end{equation}
be a discrete estimator of $Q$, which has mass at $\sigma_j$ with probability $w_j$ for $j=1,...,m$. Then, (\ref{f1}) becomes
\begin{equation}
\tilde{f}_h(x;\bw,\bsigma)=\frac{1}{2}\sum_{j=1}^mw_j\{K_h(x-\sigma_j)+K_h(x+\sigma_j)\},
\label{f2}
\end{equation}
where $\bw=(w_1,...,w_m)^\top$, and $\bsigma=(\sigma_1,...,\sigma_m)^\top$. Note that the estimator of $f$ in Bordes et al. (2007) is actually a special case of (\ref{f2}). Replacing the unknown density $f$ by (\ref{f2}), model (\ref{mmixsym}) becomes
\begin{equation}
\tilde{g}_h(x;\btheta,\bw,\bsigma)=\sum_{c=1}^C\pi_c\tilde{f}_h(x-\mu_c;\bw,\bsigma).
\label{e2.16}
\end{equation}
The estimation of $\btheta$ and $f$ is now the estimation of $\btheta, \bw$, $\bsigma$ and $m$. The log-likelihood based on (\ref{e2.16}) is then maximized by algorithms proposed in Wang (2010).

Xiang et al. (2016) studied a method that is somehow similar to Chee and Wang (2013). Instead of (\ref{f1}), they assumed the unknown density $f$ to be
\begin{equation}
\check{f}(x;Q)=\int_{\mathbb{R}^+}\frac{1}{\sigma}\phi\left(\frac{x}{\sigma}\right)dQ(\sigma),
\label{f3}
\end{equation}
where $\phi(x)$ is the standard normal density. Similar to Chee and Wang (2013), $Q$ in (\ref{f3}) is estimated by (\ref{e2.17}). Then, Xiang et al. (2016) proposed to estimate iteratively between the following two steps: the estimating of $Q$ at a $\btheta$ value, that is the estimating of $\bw,\bsigma$ and $m$ through a gradient based algorithm; and the estimating of $\btheta$ given a $\hat{Q}_n$, which is done by a regular EM algorithm. Xiang et al. (2016) argued that (\ref{f3}) includes a rich class of continuous distributions, and the resulting estimators are robust against outliers. In addition, this method avoids the selection of tuning parameters, which has always been a difficult topic.

Wu et al. (2017) proposed to estimate (\ref{e1.4}) by minimizing a profile Hellinger distance between the assumed semiparametric two-component location-shifted mixture model and a nonparametric kernel density estimator.


\subsection{With shape constraints}

To relax the parametric assumption of model (\ref{mmixture}), nonparametric shape constraints are becoming increasingly popular.

Chang and Walther (2007) proposed mixtures of log-concave distributions for clustering. Since log-concave density includes most of the common parametric distributions, such as normal, Laplace, logistic, as well as gamma and beta with certain parameter constraints, and its estimator does not need to select any tuning parameter, this method has become increasingly popular. In the meantime, since the methodology is not restricted to a parametric model, the results will not suffer from parametric model misspecification. However, without symmetry, a main drawback of such a model is that it suffers from the fundamental question of being non-identifiably. Specifically, Chang and Walther (2007) assumed that each component in (\ref{mmixture}) is log-concave, i,e, $\log f_c(x)$ is a concave function. Note that the log-likelihood of (\ref{mmixture}) assuming a log-concave $f$ is a concave function, and thus guarantees the existence of MLE.
The algorithm starts by computing the MLE of a Gaussian mixture through the regular EM algorithm. Define $\hat{\pi}_c$ and $\hat{f}_c$ as the MLE's, and
\begin{equation}
p_{ic}=\frac{\hat{\pi}_c\hat{f}_c(X_i)}{\sum_{c'=1}^C\hat{\pi}_{c'}\hat{f}_{c'}(X_i)}
\label{e2}
\end{equation}
as the classification probability of the $i$-th observation belonging to the $c$-th component. Then, in the second part of the algorithm, the E-step was the same as (\ref{e2}), where $\hat{f}_c(\cdot)$ was replaced by the log-concave MLE. The computation for $\hat{\pi}_c$ in the M-step was still $\hat{\pi}_c=\sum_{i=1}^n p_{ic}/n$, and $p_{ic}$ was used as weights for $X_i$ when the log-concave MLE $\hat{f}_c$ was computed using the methods developed in Walther (2002) and Rufibach (2007). Simulation showed that only five iterations were required in the second part of the algorithm. The model is then extended to the \emph{multivariate} situation. Assume $(N_1,\ldots,N_d)$ to be a multivariate normal distribution with mean $\b0$ and covariance matrix $\Sigma$, and $F_1,\ldots,F_d$ be cdfs of arbitrary univariate log-concave distributions. Then, within a component, observations $(X_{i1},\ldots,X_{id})^\top\in\mathbb{R}^d$ is assumed to have density $(F_1^{-1}\Phi(N_1),\ldots,F_d^{-1}\Phi(N_d))$, where $\Phi$ stands the cdf of standard normal.
Therefore, the joint density distribution for the $c$-th component is then defined as
\[f_c(x_1,\ldots,x_d)=\phi_{\b0,\Sigma}\{\Phi^{-1}F_1(x_1),\ldots,\Phi^{-1}F_d(x_d)\}\prod_{j=1}^d\frac{f_j(x_j)}{\phi_{\b0,\bI}\{\Phi^{-1}F_j(x_j)\}},\]
where $\phi_{\bmu,\Sigma}$ is the multivariate normal density with mean $\bmu$ and covariance $\Sigma$.
The resulting EM algorithm is quite similar to the univariate case, and thus is omitted here.

Hu et al. (2016) proposed the log-concave maximum likelihood estimator (LCMLE) to estimate mixture densities and provided the theoretical justification of their estimator. It was assumed that $(X_1,\ldots,X_n)$ were independent $d$-dimensional random variables with mixture distribution belonging to
\begin{equation}
\mathcal{G}_\eta=\{g: g(x)=\sum_{c=1}^C \pi_c\exp\{\phi_c(x)\} \},
\label{mixlogcon}
\end{equation}
where $\phi=(\phi_1,\ldots,\phi_C)\in \Phi_\eta$ and $\Phi_\eta=\{(\phi_1,\ldots, \phi_C): \phi_c$ is concave$, |S(\phi)|\geq \eta>0\}$ for some $\eta\in (0,1]$. Here $M_c(\phi)=\max_{x\in\mathbb{R}^d}\{\phi_c(x)\}$, $M_{(1)}(\phi)=\min_{c}\{M_c(\phi)\}$, and $M_{(C)}(\phi)=\max_{c}\{M_c(\phi)\}$, and $S(\phi)=M_{(1)}(\phi)/M_{(C)}(\phi)$. The LCMLE is then defined as
\[g_n=\arg\max_{g\in\mathcal{G}_\eta} \int \log(g)dQ_n,\]
where $Q_n$ is the empirical distribution of $(X_1,\ldots,X_n)$. Hu et al. (2016) proved the existence of the LCMLE for log-concave mixture models, and the consistency of the estimated mixture density.

Balabdaoui and Doss (2018) discussed estimation and inference for mixture of log-concave distribution (under symmetry). When testing for presence of mixing, the numerical results for the asymptotic power also show better performance of the symmetric log-concave MLE when compared to Gaussian one. Assuming that these conditions hold, the fast rate of convergence of $\hat{\theta}_n$ guarantees convergence of the nonparametric log-concave MLE of Balabdaoui and Doss (2018) to the true symmetric density at the (usual) $n^{2/5}$-rate in the $L_1$ distance.

Al Mohamad and Boumahdaf (2018) considered a semiparametric two-component mixture model when one component is parametric and the other is defined based on linear constraints on its distribution function. A new estimation method is proposed, which incorporates a prior linear information about the distribution of the unknown component and is based on $\phi$-divergences. By adding moments constraints, this method shows better performance than existing methods, which do not consider any prior information, when the proportion of the parametric component is very low.

\subsection{$d>1$}
When multivariate covariates $\bx\in\mathbb{R}^d$ $(d>1)$ are considered, a common restriction placed on $f_c$, is that each joint density $f_c$ is equal to the product of its marginal densities. In other words, the coordinates of the $\bx$ vector are independent, conditionally on the subpopulation or component from which $\bx$ is drawn. Therefore, model (\ref{mmixture}) becomes
\begin{equation}
g(\bx)=\sum_{c=1}^C\pi_c\prod_{j=1}^d f_{cj}(x_{j}),
\label{mulmodel2}
\end{equation}
Hall and Zhou (2003) showed that when $C=2$ and $d>2$, identifiability of model (\ref{mulmodel2}) can be achieved in general case. Allman et al. (2009) proved that if the density functions $f_{1k}, \ldots, f_{Ck}$ are linearly independent except possibly on a set of Lebesgue measure zero, the parameters in (\ref{mulmodel2}) are identifiable whenever $d>2$.

Benaglia et al. (2009) considered a more general case of (\ref{mulmodel2}). Assuming the coordinates of $\bx$ to be conditionally independent and that there are blocks of coordinates that have identical densities, Benaglia et al. (2009) proposed an EM-like estimation method. If all the blocks are of size 1, such as the setting in the model (\ref{mulmodel2}), then the coordinates in $\bx_i$ are conditional independent, but their distributions are all different. If there only exists one block, then the coordinates are not only conditionally independent but identically distributed, i.e., $f_{c1}(\cdot)=\cdots=f_{cd}(\cdot)$. To describe briefly their method, let $b_j$ denote the block to which the $j$-th coordinate belongs, where $1 \leq b_j \leq B$ and $B$ is the total number of such blocks. Then, model (\ref{mmixture}) becomes
\begin{equation}
g(\bx)=\sum_{c=1}^C\pi_c\prod_{j=1}^d f_{cb_j}(x_{j}).
\label{e1.5}
\end{equation}
Note that model (\ref{mulmodel2}) is a special case of the model (\ref{e1.5}) if $b_k$'s are different for every $j$'s. At the $t$-th iteration ($t=1,2,\ldots$), in the E-step, the ``posterior'' probabilities of component inclusion $p^{(t)}_{ic}$, conditional on the current estimators, is calculated in the same sense as any regular EM algorithm. In the M-step, the algorithm updates the mixing proportion by $\pi^{(t+1)}_c=n^{-1}\sum_{i=1}^n p^{(t)}_{ic}$, and the density as
\[f^{(t+1)}_{cl}(u)=\frac{1}{nhC_l \pi^{(t+1)}_c}\sum_{j=1}^d\sum_{i=1}^n p^{(t)}_{ic}I\{b_j=l\}K\left(\frac{u-x_{ij}}{h}\right),\]
for $c=1, \ldots, C,\hskip 0.1cm l=1, \ldots, B$, where $C_l=\sum_{j=1}^d I\{b_j=l\}$ is the number of coordinates in the $l$-th block. However, the authors do not discuss the theoretical properties of the estimators, nor show the ascent property of the algorithm, attained by standard EM algorithms.

To improve the work of Benaglia et al. (2009), Levine et al. (2011) introduced a smoothed log-likelihood function by replacing the component density function $f_c(x)$ by a nonlinear smoother $\mathcal{N}f_c(\bx)=\exp \int K^d_h(\bx-\bu)\log f_c(\bu)d\bu$, where $K^d_h(\bu)=h^{-d}\prod_{j=1}^dK^d(u_j/h), K^d(u)=\prod_{j=1}^dK(u_j), \bu=(u_1,\ldots,u_d)^\top$. Then the new EM algorithm, more like an maximization-minimization (MM) algorithm convergence, is proved to have the monotonicity property. 

Chauveau et al$.$ (2015) described and extended an algorithm to estimate the parameters in nonparametric multivariate finite mixture models assuming the conditional independence. Similar to the work of Benaglia et al. (2009) on model (\ref{e1.5}), Chauveau et al. (2015) also assumed that groups of conditionally iid coordinates belong to the same block. The algorithm is very close to the one of Benaglia et al. (2009), with the main difference lying in the second Minimization step
\[f^{(t+1)}_{cl}(u)=\frac{1}{nh_{cl}\pi^{(t+1)}_cC_l}\sum_{j=1}^d\sum_{i=1}^np^{(t)}_{ic}I\{b_j=l\}K\left(\frac{u-x_{ij}}{h_{cl}}\right),\]
for $c=1,\ldots,C, l=1,\ldots, B$. The algorithm is proved to attain the ascent property of a typical EM algorithm, and due to its good property and ease of calculation, the authors further extend it to the univariate models (\ref{mmixsym}).

\subsection{$d=1,C=2$ with a known component}

Consider a two-component mixture model with one known component
\begin{equation}
g(x) = (1-\pi) f_0(x) + \pi f(x-\mu), x\in\mathbb{R}
\label{e1.2}
\end{equation}
where $f_0$ is a known pdf, the pdf $f\in\mathscr{F}$, where $\mathscr{F}=\{f: f\ge 0,\int f(x)dx=1\text{ and } f(-x)=f(x)\}$, and the unknown parameters are $\btheta=(\pi,\mu)^\top$. Model (\ref{e1.2}) is motivated by the detection of differentially expressed genes under two or more conditions in microarray data analysis (Bordes et al., 2006a), and sequential clustering algorithm (Song et al., 2010). Model (\ref{e1.2}) is also quite commonly used in contamination problem, commonly seen in astronomy, biology, among other fields (Patra and Sen, 2016). It is an extension of the classical two-component mixture models in the sense that one component is supposed to be symmetric only, without assuming that it belongs to a known parametric family. In the parametric setup this model
is sometimes referred to as a contamination model.

Bordes et al. (2006a) showed the identifiability of model (\ref{e1.2}) when $f$ has third-order moment and is zero-symmetric. Similarly to Bordes et al. (2016b), the inversion of the cdf of model (\ref{e1.2}) leads to
\begin{equation}
F(x)=\frac{1}{\pi}\{G(x+\mu)-(1-\pi) F_0(x+\mu)\},
\label{e2.6}
\end{equation}
where $F_0$ is the cdf of $f_0$. Define
\begin{eqnarray*}
H_1(x;\mu,m,G)&=&\frac{\mu}{m}G(x+\mu)+\frac{m-\mu}{m}F_0(x+\mu),\\
H_2(x;\mu,m,G)&=&1-\frac{\mu}{m}G(\mu-x)+\frac{\mu-m}{m}F_0(\mu-x),
\end{eqnarray*}
where $m$ is the first-order moment of $G$. Then, by the symmetry of $F$, the estimator of $\mu$ is defined as $\hat{\mu}_n=\arg\min_\mu d\{H_1(\cdot;\mu,\hat{m}_n,\hat{G}_n),H_2(\cdot;\mu,\hat{m}_n,\hat{G}_n)\}$, where $d$ is the $L_q$ distance, $\hat{G}_n$ and $\hat{m}_n$ are the empirical versions of $G$ and $m$, derived from the a sample of size $n$. Then, $\hat{\pi}_n=\hat{m}_n/\hat{\mu}_n$. However, the estimator was shown to be numerically unstable and theoretical properties of it was not shown.

Similarly to Bordes et al., (2006a), Bordes and Vandekerkhove (2010) also considered (\ref{e2.6}), and defined
\begin{eqnarray*}
H_1(x;\btheta,G)&=&\frac{1}{\pi}G(x+\mu)+\frac{1-\pi}{\pi}F_0(x+\mu),\\
H_2(x;\btheta,G)&=&1-\frac{1}{\pi}G(\mu-x)+\frac{1-\pi}{\pi}F_0(\mu-x).
\end{eqnarray*}
However, instead of $L_q$-norm,
Bordes and Vandekerkhove (2010) considered
\[
d(\btheta)=\int_\mathbb{R}H^2(x;\btheta,G)dG(x).
\]
where $H(x;\btheta,G)=H_1(x;\btheta,G)-H_2(x;\btheta,G)$.
In order to estimate $\btheta$ by a differentiable optimization routine, another empirical version of $d$ is defined as
\[d_n(\btheta)=\frac{1}{n}\sum_{i=1}^nH^2(X_i;\btheta,\tilde{G}_n),\]
where $\tilde{G}_n(x)=\int_{-\infty}^x\hat{g}_n(t)dt$ is a smoothed version of $\hat{G}_n$, and $\hat{g}_n(x)=\frac{1}{nh}\sum_{i=1}^nK\left(\frac{x-X_i}{h}\right)$. By these improvements, the authors showed the asymptotic normality of the estimators.

Maiboroda and Sugakova (2011) considered a generalized estimating equations (GEE) method to estimate the Euclidean parameters of model (\ref{e1.2}). Let $(X_1,\ldots,X_n)$ be a sample generated from (\ref{e1.2}), and $z$, $z_0$ and $\delta$ be three random variables such that $z\sim f, z_0\sim f_0$ and $\delta\sim \rm{B}(1,\pi)$. Then, $X_i\sim \delta(z+\mu)+(1-\delta)z_0$. Denote $h_j (j=1,2)$ as two odd functions, and let $H_j(\mu)={\rm E} h_j(z_0-\mu)$ for any $\mu\in\mathbb{R}$. It is easy to see that ${\rm E} h_j(X_i-\mu)=\pi{\rm E}h_j(z)+(1-\pi)H_j(\mu)=(1-\pi)H_j(\mu)$ where second equality is derived directly by the oddness of $h_i$ and the symmetry of $f$. Motivated by this, Maiboroda and Sugakova (2011) proposed the following unbiased estimating equations for the estimation of $\btheta$:
\begin{equation*}
\left\{\begin{matrix}
\hat{h}_1(\mu)-(1-\pi)H_1(\mu)=0,  \\
\hat{h}_2(\mu)-(1-\pi)H_2(\mu)=0,
\end{matrix}\right.
\end{equation*}
where $\hat{h}_j(\mu)=n^{-1}\sum_{i=1}^nh_j(X_i-\mu)$. Maiboroda and Sugakova (2011) proved, under mild conditions, the consistency and asymptotic normality of their estimators.


Patra and Sen (2016) also studied model (\ref{e1.2}) but without assuming the symmetry of $f$. The article used ideas from shape restricted function estimation and developed ``tuning parameter free'' estimators that were easily implemented and had good finite sample performance. Consider the first estimator of the unknown cdf $F$,
\[\hat{F}(x;\pi)=\frac{\hat{G}_n(x)-(1-\pi)F_0(x)}{\pi},\]
where $\hat{G}_n$ is the empirical cdf. This estimator is easy to calculate but is not guaranteed to satisfy the conditions of a distribution function: lying between 0 and 1 and non-decreasing. Therefore, a second estimator of $F$ is proposed as $\tilde{F}(x;\pi)$, which is the minimizer of $\frac{1}{n}\sum_{i=1}^n\{W(X_i)-\hat{F}(X_i;\pi)\}^2$ over all density functions $W$.
Since the two estimators indeed all depend on $\pi$, the authors then suggest estimating $\pi$ by
\[\hat{\pi}=\inf\left\{p\in(0,1): p d_n\{\hat{F}(x;p),\tilde{F}(x;p)\}\leq \frac{c_n}{\sqrt{n}}\right\},\]
where $c_n$ is a sequence of constants and $d_n$ stands for the $L_2$ distance. It is shown that for a broad range of $c_n$, the estimating procedure is proved to be consistent. Further study shows that the ``elbow'' of $pd_n(\hat{F}(x;p),\tilde{F}(x;p))$, i.e., the point that has the maximum curvature, is a good estimator of $\pi$, and is free of tuning. Once an estimator of $\pi$ is decided, say $\hat{\pi}_n$, then it is natural to estimate $F$ by $\tilde{F}(\hat{\pi}_n)$.

There are some generalization of model (\ref{e1.2}). For example, Hohmann and Holzmann (2013) studied a generalization of (\ref{e1.2}),
\[
g(x) = (1-\pi) f_0(x-\nu) + \pi f(x-\mu), x\in\mathbb{R},
\]
where $\nu$ is another non-null location parameter. They showed identifiability under assumptions on the tails of the characteristic function for the true underlying mixture, and also constructed asymptotically normal estimators, with methodologies quite similarly to Bordes and Vandekerkhove (2010).

Xiang et al. (2014) and Ma and Yao (2015) studied another transformation of model (\ref{e1.2}), by assuming $f_0$ to be known but with a unknown parameter. That is,
\begin{equation}
g(x;\btheta,f)= (1-\pi) f_0(x;\xi) + \pi f(x-\mu), x\in\mathbb{R},
\label{semimix}
\end{equation}
where $\xi$ is an unknown parameter, and $\btheta=(\pi,\mu,\xi)^\top$ is the vector of unknown parameters. Ma and Yao (2015) studied the identifiability conditions of model (\ref{semimix}) and proposed a general class of estimation equations based estimators, which also had a nice connection to the most efficient estimator for all parameters in the sense of semiparametric efficiency. The estimator proposed by Xiang et al. (2014) is based on the minimum profile Hellinger distance, and its theoretical properties are investigated. Define the Hellinger distance between two functions $g_1, g_2$ as
\[d_H(g_1,g_2)=\|g_1^{1/2}-g_2^{1/2}\|,\]
where $\|\cdot\|$ denotes the $L_2$-norm, and it is a natural idea to estimate $\btheta$ and $f$ by minimizing $d_H\{g(\cdot;\btheta,f),\hat{g}_n\}$ over $\btheta\in\Theta$ and $f\in\mathcal{F}$, where $\hat{g}_n$ is a nonparametric kernel density estimator of the data. Note that this optimization problem involves both the parametric part $\btheta$ and the nonparametric part $f$, and so the authors suggested to apply the profile idea to implement the calculation. First, for any $\btheta$, define $f(\btheta,\hat{g}_n)=\arg\min_{l\in\mathcal{F}}d_H\{g(\cdot;\btheta,l),\hat{g}_n\}$, and then the minimum profile Hellinger distance estimator of $\btheta$ is defined as $\hat{\btheta}_H=\arg\min_{\btheta\in\Theta}d_H[g\{\cdot;\btheta,f(\btheta,\hat{g}_n)\},\hat{g}_n]$. The algorithm works by iterating between updating the parameter $\btheta$ and updating the nonparametric function $f$. Xiang et al. (2014) further showed the asymptotic normality of the minimum profile Hellinger distance estimator.

Assuming the parametric component $f_0$ in (\ref{semimix}) follows a Pareto distribution with unknown parameters $\xi$, Huang et al. (2018) proposed another special case of model (\ref{semimix}). The identifiability is discussed, and a novel estimation method is studied using smoothed likelihood and profile-likelihood techniques. A smoothing kernel $K_{h,\mu}(x,t)=(2h)^{-1}[K\{(x-t)/h\}+K\{(2\mu-x-t)/h\}]$ is defined, which is $\mu$-symmetric, and correspondingly a nonlinear smoothing operator for $f(\cdot)$ is defined as
\[\mathcal{N}_\mu f(x)=\exp \left\{\int K_{h,\mu}(x,t)\log f(t)dt\right\}.\]
Replacing $f$ by its nonlinear smoother, the smoothed log-likelihood of a data is then
\[\ell(\mu,\pi,\xi,f)=\sum_{i=1}^n \log\{(1-\pi)f_0(X_i;\xi)+\pi \mathcal{N}_\mu f(X_i)\}.\]
The authors proposed an estimation method that separates $\mu$ from $\pi, \xi$ and $f$. Given a known $\mu$, or an estimator of it, denoted by $\mu_0$, then, the maximum likelihood estimator of $\pi, \xi$ and $f$ can be calculated by maximizing $\ell(\mu_0,\pi,\xi,f)$, via an EM algorithm. Denote the estimators by $\hat{\pi}_\mu, \hat{\xi}_\mu$, and $\hat{f}_\mu(\cdot)$. Then, the estimator of $\mu$ is through maximizing the profile likelihood $\hat{\ell}_p(\mu)=\ell(\mu,\hat{\pi}_\mu,\hat{\xi}_\mu,\hat{f}_\mu),$ which can be done through some advanced numerical methods.

Nguyen and Matias (2014) studied a special case of (\ref{e1.2}) when $f_0(\cdot)=1$, and proved an impossibility result. They showed that the quadratic risk of any estimator of $\pi$ does not have a parametric convergence rate when $f$ was not $0$ on any non-empty interval. This happens mainly because  the Fisher information for the model was $0$ when $f$ was bounded away from $0$ for all non-empty interval. We conjecture that such results might also hold for the two symmetric component density case and the multivariate case, and this could be an interesting topic for future work.




\section{Semiparametric mixture of regressions}
\subsection{Introduction}

In a typical finite mixture of regressions (FMR) model, assume $\{(\bx_i,y_i), i=1,\ldots,n\}$ is a random sample from the population $(\bx, Y)$, where $\bx_i=(x_{i,1}, \ldots , x_{i,p})^\top$ for $p < n$ is a vector of predictors. The goal is to describe the conditional distribution of $Y_i|\bx_i$ using a mixture of linear regressions with assumed Gaussian errors. That is, let $\mathcal{C}$ be a latent class index random variable with $P(\mathcal{C}=c|\bx)=\pi_c$ for $c=1,\ldots,C$, given $\{\mathcal{C}=c\}$, suppose that the response $y$ depends on $\bx$ in a linear way $y=\bx^\top\bbeta_c+\varepsilon_c$, where $\varepsilon_c\sim N(0,\sigma_c^2)$. Then, the conditional distribution of $Y$ given $\bx$ is
\begin{equation}
Y|{\bx}\sim\sum_{c=1}^C\pi_c\phi(y|\bx^\top\bbeta_c,\sigma_c^2),
\label{fmr}
\end{equation}
where $\btheta=(\pi_1,\ldots,\pi_C,\bbeta_1,\ldots,\bbeta_C,\sigma_1^2,\ldots,\sigma_C^2)^\top$ is the vector of parameters, $\phi(y|\mu,\sigma^2)$ is the normal density with mean $\mu$ and variance $\sigma^2$, $0\leq\pi_c\leq 1$, and $\sum_{c=1}^C \pi_c=1$, see McLachlan and Peel (2000) for comprehensive discussions.

\subsection{Mixture of regression models with varying proportions}
In a parametric mixture of regressions model, the mixing proportions are assumed to be known and fixed as $\pi_c$, $c=1,\ldots,C$. However, if the covariates $\bx$ contain some information about the relative weights, then, model (\ref{fmr}) might not be efficient enough. In the following, several FMR models with varying proportions are discussed. The error density is assumed to be known throughout the section.

The first model is
\begin{equation}
Y|{\bx}\sim\sum_{c=1}^C\pi_c(\bx)\phi(y|\bx^\top\bbeta_c,\sigma_c^2)
\label{e2.1}
\end{equation}
which identifiability was discussed by Huang and Yao (2012) under some mild conditions.
If $\pi_c(\bx)$ is modeled as a logistic function, then model (\ref{e2.1}) becomes the hierarchical mixtures of experts (HME, Jacobs et al., 1997) in neural network. Young and Hunter (2010), on the other hand, modeled $\pi_c(\bx)$ as
\begin{equation}
\pi_c(\bx_i)=E[z_{ic}|\bx_i],
\label{cla}
\end{equation}
where $z_{ic}$ is a component indicator variable that is 1 if the $i$-th observation is from the $c$-th component, and 0 otherwise. Note that if one treats $z_{ic}$ as a response, then (\ref{cla}) indicates nothing but a mean structure in a regression analysis. In that setup, Young and Hunter (2010) proposed to estimate $\pi_c(\bx_i)$ by local polynomial regression (Fan and Gijbels, 1996) as
\begin{equation}
\arg\min_\balpha \sum_{l=1}^n K_\bh(\bx_i-\bx_l)\left\{z_{ic}-\left ( \alpha_0+\sum_{t=1}^p\alpha_{t}(x_{i,t}-x_{l,t})\right )\right\}^2,
\label{e2.13}
\end{equation}
where $\balpha=(\alpha_0,\alpha_1,\ldots,\alpha_p)^\top$, and $K_\bh(\bx_i-\bx_l)$ is a multivariate kernel density function. However, since $z_{ic}$ is not known in reality, they proposed to run the EM algorithm for model (\ref{fmr}) first, and then used the converged value of the classification probability in E-step, denoted by $p_{ic}^\infty$, to replace for $z_{ic}$. Given estimators of $\pi_c(\bx)$ for $c=1,\ldots,C$, $\bbeta_c$ and $\sigma_c$ can then be estimated through a regular EM algorithm. However, due to the ``curse of dimensionality'', the authors only did simulation study for $p=1$ case, and argued that extra caution should be given for high-dimensional predictors cases. Theoretical results were not discussed for this method.

Huang and Yao (2012), on the other hand, studied $\pi(\bx)$ fully nonparametrically, and proposed a one-step backfitting procedure to achieve the optimal convergence rates for both regression parameters and the nonparametric functions of mixing proportions. They further derived the asymptotic bias and variance of the one-step estimator.

Allowing the response to come from other distribution families, Model (\ref{e2.1}) can then be extended to mixture of GLMs with varying proportions (Wang et al. 2014):
\begin{equation}
Y|\bx\sim\sum_{c=1}^C\pi_c(\bx)f_c(y|\bx,\btheta_c),
\label{e2.15}
\end{equation}
where $f_c$ is a function of the exponential family, the mean of which is given by $\mu_c(\bx)=g_c^{-1}(\bx^\top\bbeta_c)$, where $g_c(\cdot)$ is a component specific link function. For example, when a binomial response $Y$ is considered, Cao and Yao (2012) studied a special case of (\ref{e2.15}), where both the component proportions and the success probabilities depend on the predictors nonparametrically. That is,
\begin{equation}
Y|_{X=x}\sim \pi_1(x)\text{Bin}(y;N,0)+\pi_2(x)\text{Bin}\{y;N,p(x)\},
\label{binom}
\end{equation}
where $\pi_1(x)+\pi_2(x)=1$, and Bin$(Y;N,p)$ denotes the pmf of a binomial distribution random variable $Y$ with number of trials $N$ and success probability $p$. Note that the first component is a degenerate distribution with mass 1 on 0, and therefore, model (\ref{binom}) has wide application in data with extra number of zeros. Cao and Yao (2012) successfully applied model (\ref{binom}) to a rain data
from a global climate model and a historical rain data from Edmonton, Canada.

\subsection{Nonparametric errors}
Traditional FMR models (\ref{fmr}) are all based on the assumption of normally distributed errors. The estimation results might be biased or even misleading if this assumption is problematic. In the following, we review several methodologies that relax this condition.

Hunter and Young (2012) studied a FMR model where linearity is still assumed within each component, but instead of normality, the error terms were modeled fully nonparametrically as $\varepsilon_i\sim g$. That is,
\begin{equation}
Y|{\bx}\sim\sum_{c=1}^C\pi_cg(y-\bx^\top\bbeta_c).
\label{e2.3}
\end{equation}
Without loss of generality, $g$ is assumed to have median 0. The identifiability of model (\ref{e2.3}) can be achieved whenever the regression planes are not parallel. If some further conditions are to put on $g$, then (\ref{e2.3}) can still be identifiable even when the regression planes are parallel. Similar to Levine et al. (2011), Hunter and Young (2012) proposed an estimation procedure that maximizes the following smoothed log-likelihood
\begin{equation*}
\ell_s(\bpi,\bbeta,g)=\sum_{i=1}^n\log\left\{\sum_{c=1}^C\pi_c \mathcal{N}_h g(y_i-\bx_i^\top\bbeta_c)\right\}.
\label{e2.4}
\end{equation*}
where $\mathcal{N}_h g=\exp\int h^{-1}K\left\{(x-u)/h\right\}\log g(u)du$ is a nonlinear smoother. The effectiveness of the new methods was demonstrated through numerical studies. Ma et al. (2018) extended the identifiability result for the model (\ref{e2.3}) by allowing different component error densities and further established the consistency and asymptotic normality of their proposed estimators as well as the estimator proposed by Hunter and Young (2012).


Hu et al. (2017) assumed the error densities to be log-concave. That is, the model has the same form as (\ref{e2.3}), whereas $g_c(x)=\exp\{\phi_c(x)\}$ for some unknown concave function $\phi_c(x)$.

The two methods studied above all focus on the mean regression. By regressing the conditional quantiles (such as median) on the covariates without any parametric assumption, Wu and Yao (2016) studied a semiparametric mixture of quantile regressions model. Given $\mathcal{C}=c$,
\begin{equation}
Y=\bx^\top\bbeta_c(\tau)+\varepsilon_c(\tau),
\label{e10.1}
\end{equation}
where $\bbeta_c(\tau)=(\beta_{0c}(\tau),\ldots,\beta_{pc}(\tau))^\top$ is the $\tau$-th quantile regression coefficient for the $c$-th component. The only assumption on the error density $g_c(\cdot)$ is that the $\tau$-th quantiles are zero. Model (\ref{e10.1}) is believed to be more robust than regular FMR model, and could reveal more detailed data structure. Wu and Yao (2016) proposed an EM-type algorithm by incorporating the kernel regression to estimating the parameters and error densities.

\subsection{Semiparametric mixtures of nonparametric regressions}
In the traditional FMR model (\ref{fmr}) and the models discussed above, linearity is always assumed in the mean functions. In the following, different models were proposed to relax this assumption.

Motivated by a US house price index data, Huang et al. (2013) proposed the following model:
\begin{equation}
Y|_{X=x}\sim \sum_{c=1}^C \pi_c(x)N\{m_c(x),\sigma_c^2(x)\},x\in\mathbb{R},
\label{e6.1}
\end{equation}
where $\pi_c(\cdot), m_c(\cdot), \sigma^2_c(\cdot)$ are unknown but smooth functions, and $\sum_{c=1}^C \pi_c(\cdot)=1$. Note that the errors are assumed to follow a normal distribution, and so model (\ref{e6.1}) is still considered a semiparametric mixture model. Since there are nonparametric functions, kernel regression is used in a modified EM algorithm. Specifically, like any regular EM algorithm, at $(t+1)$-th iteration $(t=1,2,\ldots)$, a ``posterior'' probability is calculated and labeled as $p_{ic}^{(t+1)}$, based on current estimators. Then, at the M-step, the following local objective function with respect to $\pi_c, m_c$ and $\sigma_c$ is maximized to update the estimators
\[\sum_{i=1}^n\sum_{c=1}^Cp_{ic}^{(t+1)}[\log \pi+\log\phi\{Y_i|m_c,\sigma_c^2\}]K_h(X_i-x).\]


Model (\ref{e6.1}) is flexible enough, but also suffers from a lack of efficiency as a side effect. Taking into account both matters, Xiang and Yao (2016) suggested a new model by assuming the mixing proportions and variances to be constant. The model is defined as:
\begin{equation}
Y|_{X=x}\sim \sum_{c=1}^C \pi_c\phi\{y|m_c(x),\sigma_c^2\},
\label{e12.1}
\end{equation}
where $m_c(\cdot)$s are unknown smooth functions. Due to the co-existence of both global and local parameters, model (\ref{e12.1}) is more difficult to estimate. An efficient one-step backfitting estimation procedure, similar to the ones discussed in Huang and Yao (2012) and Cao and Yao (2012), was proposed. A generalized likelihood ratio test was also proposed to compare between model (\ref{e6.1}) and model (\ref{e12.1}), and was shown to have the Wilks types of results.

Similarly to the issue discussed in the previous section, due to the application of kernel regression in the estimation procedure, model (\ref{e6.1}) and model (\ref{e12.1}) are not suitable for data with high dimensional predictors. As a result, Xiang and Yao (2017) studied a series of FMR model with single-index. First, replacing the one-dimensional covariate $x$ in (\ref{e6.1}) by $\balpha^\top\bx$, a mixture of single-index models (MSIM) is defined as:
\begin{equation}
Y|{\bx}\sim\sum_{c=1}^C\pi_c(\balpha^\top\bx)\phi\{y|m_c(\balpha^\top\bx),\sigma_c^2(\balpha^\top\bx)\}.
\label{e13.1}
\end{equation}
When $C=1$, model (\ref{e13.1}) reduces to a single index model (Ichimura, 1993; H$\ddot{a}$rdle et al., 1993). If $\bx$ is a scalar, then model (\ref{e13.1}) reduces to model (\ref{e6.1}). 
Models with nonparametric means are flexible enough, but hard to estimate and difficult to interpret. As a result, introducing single-index into the mixing proportions of model (\ref{e2.1}), Xiang and Yao (2017) proposed the following model:
\begin{equation}
Y|{\bx}\sim\sum_{c=1}^C\pi_c(\balpha^\top\bx)N(\bx^\top\bbeta_c,\sigma_c^2).
\label{e6.2}
\end{equation}
The global parameters $\balpha,\bbeta=(\bbeta_1^\top,\ldots,\bbeta_C^\top)^\top$, $\bsigma^2=(\sigma_1^2,\ldots,\sigma_C^2)^\top$ and the nonparametric functions $\bpi(\cdot)=(\pi_1(\cdot),\ldots,\pi_C(\cdot))^\top$ are estimated alternately by fixing the others.

\subsection{Semiparametric regression models for longitudinal/functional data}
In the following, we introduce the applications of semiparametric mixture models to more complex data. Early works of such mixture model can be found in Yao et al. (2011), which extended traditional functional linear models to the framework of classical mixture regression models, and proposed a functional mixture regression model.

Rich in information, intensive longitudinal data (ILD) are becoming increasingly popular in behavioral sciences. However, since ILD are always heterogeneous and nonlinear, they are difficult to analyze. Dziak et al. (2015) proposed a mixture of time-varying effect models (MixTVCM), which incorporated time-varying effect model (TVEM) into a finite mixture model framework. Conditional on time-invariant subject-level covariates $s_1,\ldots,s_Q$, the probability that individual $i$ comes from class $c$ is
\begin{equation*}
\pi_{ic}=P(\mathcal{C}_i=c)=\frac{\exp\left (\gamma_{0,c}+\sum_{q=1}^Q\gamma_{1qc}s_q\right )}{\sum_{t=1}^k\exp\left (\gamma_{0,t}+\sum_{q=1}^Q\gamma_{1qt}s_q\right )},
\label{e4.1}
\end{equation*}
and within each component, the means are assumed to be the same as the TVEM model in Tan et al. (2012):
\begin{equation*}
\mu_{ij}=E(y_{ij}|\mathcal{C}_i=c)=\beta_{0c}(t_{ij})+\beta_{10}(t_{ij})x_{ij1}+\ldots+\beta_{pc}(t_{ij})x_{ijP},
\label{e4.2}
\end{equation*}
where $x_1,\ldots,x_P$ is the observation-level covariates. The covariance structure of $Y_{ij}$ is assumed to satisfy
\begin{equation*}
\text{cov} (y_{ij}, y_{ij'})=\sigma_a^2\rho^{|t_{ij}-t_{ij'}|}+\sigma_e^2,
\label{e4.3}
\end{equation*}
where $\sigma_a^2$ is the variance of a normally distributed subject-level error, and $\sigma_e^2$ is the variance of a normally distributed observation-level error.
Since normality is assumed for error distributions, even nonparametric in means, MixTVEM is still considered semiparametric. In order to make the model identifiable, it is assumed that individuals are clustered into one and only one latent class. In the presence of mixture structure, EM algorithm is used for the estimation, in which the penalized B-spline is used to approximate $\beta(\cdot)$'s, where the penalization is considered to ensure a smooth and parsimonious shape.

To deal with data collected at irregular, possibly subject-depending time points, which occurs when data are functional and inhomogenous in nature, Huang et al. (2014), proposed a new estimation procedure for the mixture of Gaussian processes. Conditional on $\mathcal{C}=c$, the model assumes
\begin{equation}
y_{ij}=\mu_c(t_{ij})+\sum_{q=1}^\infty \xi_{iqc}\nu_{qc}(t_{ij})+\varepsilon_{ij}, i=1,\ldots,n; j=1,\ldots, N_i,
\label{e5.1}
\end{equation}
where $\varepsilon_{ij}$'s are iid and $N(0,\sigma^2)$ distributed, $\mu_c(t)$ is the mean of the Gaussian process, which corresponding covariance function is $G_c(s,t)$, and $\xi_{iqc}$ and $\nu_{qc}(t)$ are the functional principal component (FPC) score and eigenfunctions of $G_c(s,t)$ (Karhunen-Lo$\grave{e}$ve theorem, Roger and Pol, 1991).

To analyze heterogeneous functional data, for each component, Wang et al. (2016) proposed to model, given $\{\mathcal{C}=c\}$, $\{y(t),t\in T\}$ in a functional-linear way:
\begin{equation}
y(t)=\bX(t)^\top\bbeta_c(t)+\varepsilon_c(t),
\label{e8.1}
\end{equation}
where $\bX(t)$ is a random covariate process of dimension $p$, and $\bbeta_c(t)$ is a smooth regression coefficient function of $c$-th component. $\varepsilon_c(t)$ is a Gaussian process with mean zero, independent of $\bX(t)$, and is assumed as
\[\varepsilon_c(t)=\zeta_c(t)+e(t),\]
where $\zeta(t)$ denotes a trajectory process with covariance $\Gamma_c(s,t)=\text{cov}\{\xi_c(s),\xi_c(t)\}$, and $e(t)$ is the measurement error with constant variance $\sigma^2$. For ease of notation, define $y_{ij}=y_i(t_{ij})$, $j=1,\ldots,N_i$, and similarly, define $\varepsilon_{cij}, e_{ij}$, etc. Similar to Huang et al. (2014), by Karhunen-Lo$\grave{e}$ve theorem, model (\ref{e8.1}) can be represented as
\begin{equation}
y_{ij}=\bX_i(t_{ij})^\top\bbeta_c(t_{ij})+\sum_{q=1}^\infty\xi_{iqc}v_{qc}(t_{ij})+e_{ij},
\end{equation}
where $v_{qc}(\cdot)$'s are eigenfunctions of $\Gamma_c(s,t)$ and $\lambda_{qc}$'s are corresponding eigenvalues, and $\xi_{iqc}$'s are uncorrelated FPC of $\zeta_c(t)$ satisfying $E(\xi_{iqc})=0$, $\text{var}(\xi_{iqc})=\lambda_{qc}$. Ignoring the correlation structure, $y_{ij}$ can be thought to be coming from the following mixture of Gaussian process
\[y(t)\sim \sum_{c=1}^C\pi_c N\{\bX(t)^\top\bbeta_c(t),\sigma_c^{*2}(t)\},\]
where $\sigma_c^{*2}(t)=\Gamma_c(t,t)+\sigma^2$. Then, the parameters $\pi_c, \bbeta_c(\cdot)$, and $\sigma_c^{*2}(\cdot)$ can be estimated by an EM-type algorithm, very close to the one discussed above in Huang et al. (2014).

\subsection{Some additional topics}
In addition to what we discussed above, there are some other interesting topics. For example, Vandekerkhove (2013) studied a two-component mixture of regressions model in which one component is entirely known while the mixing proportion, the slope, the intercept, and the error distribution of the other component are unknown. The method proposed by Vandekerkhove (2013) performs well for data sets of reasonable size, but since it is based on the optimization of a contrast function of size $O(n^2)$, the performance is not desirable as the sample size increases. Bordes et al. (2013) also studied the same model as Vandekerkhove (2013), and proposed a new method-of-moments estimator, whose order is of $O(n)$. Young (2014) extended the mixture of linear regression models to incorporate changepoints, by assuming one or more of the components are piecewise linear. Such model is a great combination of traditional mixture of linear regression models and standard changepoint regression model. Faicel (2016) proposed a new fully unsupervised algorithm to learn regression mixture models with unknown number of components. Unlike the standard EM for mixture of regressions, this method did not require accurate initialization. Montuelle and Le Pennec (2014) studied a mixture of Gaussian regressions model with logistic weights, and proposed to estimate the number of components and other parameters through a penalized maximum likelihood approach. Butucea et al. (2017) considered a non-linear mixture of regression models with one known component. A local estimation procedure based on the symmetry of the local noise is proposed to estimate the proportion and locations functions. Huang et al. (2017a) proposed a semiparametric hidden Markov model with non-parametric regression, in which the mean and variance of emission model are unknown smooth functions, see also Gassiant et al. (2017), de Castro et al. (2017), Gassiant et al. (2016), Gassiant and Rousseau (2016), and Dannemann et al. (2014) for more discussion on nonparametric/semiparametric hidden Markov models, and Gassiat (2017) for a survey of mixtures of nonparametric components and hidden Markov models. Huang et al. (2017b) established the identifiability and investigated the statistical inference for mixture of varying coefficient models, in which each mixture component follows a varying coefficient model and the mixing proportions and dispersion parameters are unknown smooth functions.

\section{Discussion}
This article summarizes several semiparametric extensions of the standard parametric mixture of locations model and regressions model. Detailed model settings and corresponding estimation methods  are presented. As we have seen, this field has received a lot of interest, but there are still a great number of questions and issues that remain to be addressed. Choosing the number of components in mixture models is an important problem, which have attracted a lot of attention in statistical research. For parametric mixture models, some popular and simple approaches are the use of the information criteria, such as AIC or BIC, and likelihood ratio tests, see McLachlan and Peel (2000), Chen et al. (2004) and Chen and Li (2009) for more details. For semiparametric mixture models, however, one main difficulty lies in the definition of model complexity. Huang et al. (2013) applied the degrees of freedom derived in Fan et al. (2001), and proposed a information criterion approach for model selection. It is still an open and interesting topic, waiting for other attempts. In addition, since lots of the models we discussed above are closely connected or even nested, then in addition to data driven methods, it is natural to develop some testing procedure to formally select the model. See, for example, Pommeret and Vandekerkhove (2018), which investigates a semiparametric testing approach to answer if the Gaussian assumption made by McLachlan et al. (2006) on the unknown component of their false discovery type mixture model was a posteriori correct or not.

Furthermore, for many of the semiparametric mixture models we reviewed, see, for example, Bordes et al. (2007), Benaglia et al. (2009), Hunter and Young (2012), only some EM-type algorithms are proposed without rigorous theoretical justifications or asymptotic properties. Due to nonparametric kernel density estimates used in those EM-type algorithms, the algorithms do not possess the ascent property of a standard EM algorithm. It requires more research to establish some theoretical properties, such as the optimal convergence rate and semiparametric efficiency, about the semiparametric mixture estimators. We hope that this review article could inspire more researchers to shine more lights on this topic.

\section*{Acknowledgements}

Xiang's research is supported by NSF of China [grant no. 11601477] and Zhejiang Provincial NSF of China [grant no. LQ16A010002]. Yao's research is supported by NSF grant DMS-1461677 and Department of Energy with the award DE-EE0007328. Yang's research was supported by the National Natural Science Foundation of China grant 11471086, the National Social Science Foundation of China grant 16BTJ032, Science and Technology Program of Guangzhou 2016201604030074 and and Science and Technology Planning Project of Guangdong 2016A050503033.

\end{document}